\documentclass[twocolumn,trackchanges]{aastex701}

\usepackage{hyperref}
\usepackage{bm}
\hypersetup{linkcolor=blue,citecolor=blue,filecolor=cyan,urlcolor=blue}

\begin{document}

\title{The effect of photon re-scattering due to cosmic reionization on 21-cm images and power spectra}

\author[orcid=0000-0001-6129-0118,sname='Sikder']{Sudipta Sikder}
%\altaffiliation{Kitt Peak National Observatory}
\affiliation{School of Physics and Astronomy, Tel-Aviv University, Tel-Aviv, 69978, Israel}
\email[show]{sudiptas@mail.tau.ac.il}

\author[orcid=0000-0002-1557-693X,gname=Rennan, sname='Barkana']{Rennan Barkana} 
%\altaffiliation{Las Campanas Observatory}
\affiliation{School of Physics and Astronomy, Tel-Aviv University, Tel-Aviv, 69978, Israel}
\email{barkana@tauex.tau.ac.il}

\author[orcid=0000-0002-1369-633X
,gname=Anastasia, sname='Fialkov']{Anastasia Fialkov} 
\affiliation{Institute of Astronomy, University of Cambridge, Madingley Road Cambridge, CB3 0HA, UK}
\affiliation{Kavli Institute for Cosmology, Madingley Road Cambridge, CB3 0HA, UK}

\email{anastasia.fialkov@gmail.com}

\begin{abstract}

The 21-cm signal from neutral hydrogen serves as a critical tool for unraveling the astrophysical processes that shaped  cosmic dawn and the epoch of reionization. We explore the usually overlooked impact of re-scattering of 21-cm photons during and after reionization, similarly to cosmic microwave background photons.
This scattering affects the observed brightness temperature by mixing the original signal with light scattered into the line of sight from other regions, effectively at the mean 21-cm brightness temperature. 
This gives a small but significant effect. 
We show that it attenuates the fluctuations in a 21-cm image by $4-7\%$, while reducing the 21-cm power spectrum by a scale-independent $7–13\%$ during cosmic dawn and reionization. Incorporating this correction is vital for precisely comparing theoretical predictions with observations from experiments such as NenuFAR, LOFAR, and HERA, and the upcoming Square Kilometre Array. 

\end{abstract}



\section{Introduction}

The Cosmic Microwave Background (CMB) radiation, a snapshot of the Universe at approximately 380,000 years after the Big Bang, originates from the epoch of recombination when the plasma transitioned from opaque to transparent, releasing photons from the surface of last scattering at $z \approx 1100$ \citep{Planck2020}. As these photons propagated to the present day ($z=0$), they remained subject to weak interactions with intervening matter, imprinting subtle distortions on the observed CMB temperature and polarization anisotropies \citep{hu2002}. These secondary effects provide valuable probes of cosmic evolution, including reionization and large-scale structure formation.

Among the scattering processes affecting CMB photons post-recombination are Thomson scattering by free electrons, and the Sunyaev--Zel'dovich (SZ) effect in hot intracluster gas \citep{Birkinshaw1999}. Thomson scattering, the  deflection of low-energy photons by free electrons without significant energy transfer, was common during the ionized early Universe and re-emerged during the epoch of reionization (EoR) at $z \sim 6 - 15$, when ultraviolet photons from the first stars and galaxies reionized the intergalactic medium \citep{BARKANA2001125}. 

During and after the EoR, when free electrons once again became abundant, they re-scattered CMB photons. A similar effect 
should operate on 21-cm photons coming from those redshifts and higher (including cosmic dawn, $z \sim 15 - 30$). As 21-cm cosmology is the most promising probe of astrophysics at those epochs, \citep[e.g.,][]{furlanetto2006, Pritchard2012, barkana18book}, precise predictions are essential.
Although only a small fraction of the photons re-scattered, this scattering mixed the intrinsic 21-cm emission from neutral hydrogen with 21-cm photons scattered into the line of sight from other directions, effectively diluting the observed brightness temperature contrast against the CMB background. This effect is normally absent from 21-cm predictions, though it has been included in predictions for the Dark Ages 21-cm signal \citep{Mondal2023}.

In this paper, we quantify the impact of the scattering between CMB photons and neutral hydrogen on the 21-cm brightness temperature and its power spectrum during the cosmic dawn and EoR. Although the effect is small, we find that the 21-cm power spectrum is suppressed by around $10\%$ for a CMB optical depth as estimated by the Planck experiment \citep{Planck2020}. It is important to note that the effect is redshift dependent, and depends both on the (currently uncertain) total optical depth and on the (even more uncertain) reionization history. 

\section{Methodology}

\subsection{Simulating the 21-cm signal}

We use our simulation code 21-cm Semi-numerical Predictions Across Cosmological Epochs \citep[21cmSPACE,][]{visbal12, fialkov14b, fialkov19, cohen17} to simulate the 21-cm signal and ionization history of the early Universe at $z\geq6$. Throughout this work, we consider two populations of stars featuring updated Population III models as well as a realistic transition from Population III to Population II stars \citep{Gessey_Jones2022,Liu2025}. The astrophysical parameters adopted in this work are summarized in Table~\ref{tab:parameters}. 

\begin{table}

\centering
\begin{tabular}{lcc} 
\hline
Parameter    & Value & Description \\ 
\hline\hline
$f_{\star, \rm{II}}$  & $0.1$ & Pop II star formation efficiency \\

$f_{\star, \rm{III}}$ & $0.01$ & Pop III star formation efficiency \\

$V_{\rm{c}}$  & $V_{\rm{c}}$ km s$^{-1}$    & Minimum circular velocity \\
$f_{\rm{X, II}}$  & $1$ & Pop II X-ray efficiency \\
$f_{\rm{X, III}}$ & $0$ & Pop III X-ray efficiency \\
%$\alpha$  & $1-1.5$      & Slope of X-ray SED  \\
%$E_{\rm{min}}$  & $0.1-3.0$ keV   & X-ray SED low energy cutoff\\
%SED & 1 &  Fragos et al. 2013 Pop II XRB X-ray spectrum\\
%$\tau$  & $0.028-0.098$     & CMB optical depth \\
$\zeta_{\rm{II}}$ & $\zeta$ & Pop II ionization efficiency \\
$\zeta_{\rm{III}}$ & 49.5 & Pop III ionization efficiency\\ 

%$f_{\rm{esc}}$ & 0.3 & Escape fraction of ionizing photons for PopIII \\

$R_{\rm{mfp}}$  & $50$ Mpc  & Mean free path for ionizing photons \\
%$f_{\rm{Radio}}$  & $0$    & Radio production efficiency\\
%Pop III  to Pop II & Intermediate (30 Myr)&Populations of stars and their transition \\

\hline
\end{tabular}
\caption{The astrophysical parameters of our 21cmSPACE code and their values. The $[V_{\rm{c}}, \zeta]$ parameters for our ``Standard" and ``High $\tau$" models are $[35.5, 10]$ and $[16.5, 3.5]$, respectively.}\label{tab:parameters}

\end{table}

The brightness temperature of the redshifted 21-cm radiation (from some direction at a redshift $z$) is given by \citep{Madau1997, Pritchard2012, barkana18book}:
\begin{equation}
    T_{21} = \frac{T_{\rm S} - T_{\rm rad}}{1+z}(1 - e^{-\tau_{21}})\ ,
\label{eq:T21}\end{equation}
where $T_{\rm{S}}$ is the spin temperature of neutral hydrogen, $T_{\rm{rad}}$ is the background radiation temperature, and $\tau_{21}$ is the 21-cm optical depth (distinct from the optical depth to Thomson scattering, which is our main focus). We make the standard assumption that the background is the cosmic microwave background (CMB), so that $T_{\rm rad} = T_{\rm CMB} = 2.725(1 + z)$ K.

The power spectrum of the 21-cm brightness temperature fluctuations is defined as
\begin{equation}
\langle \tilde{\delta}_{T_{21}}(\bm{k})\tilde{\delta}_{T_{21}}^*(\bm{k^{\prime}}) \rangle = (2\pi)^3\delta_D(\bm{k}-\bm{k^{\prime}})P_{21}(k)\ ,
\end{equation}
where $\bm k$ is the comoving wavevector, $\delta_{D}$ is the Dirac delta function, and $\tilde{\delta}_{T_{21}}(\bm{k})$ is the Fourier transform of the 21-cm fluctuation $\delta_{T_{21}}(\bm{x})$, which is defined by $\delta_{T_{21}} (\bm{x}) = (T_{21}(\bm{x}) - \langle T_{21}\rangle)/\langle T_{21}\rangle$, with angular brackets denoting the spatial average. We express the 21-cm power spectrum in terms of the squared fluctuation in units of mK$^2$:
\begin{equation}\label{eqn:PK}
    \Delta_{21}^2 = \frac{k^3}{2\pi^2}P_{21}(k) \langle T_{21}\rangle^2\ , 
\end{equation}
where $k^3P_{21}(k)/{2\pi^2}$ is the dimensionless squared fluctuation.

Thomson scattering between 21-cm photons and neutral hydrogen mixes the intrinsic 21-cm signal with scattered light from adjacent regions along the line of sight, leaving the sky-averaged (global) 21-cm signal unaffected. However, it changes 21-cm brightness temperature maps and their power spectra.

\subsection{The optical depth due to Thomson scattering}

For a given reionization history, as simulated with a run of 21cmSPACE, we compute the scattering of 21-cm photons by free electrons, integrating the contributions from ionized hydrogen and helium across cosmic time:
\begin{equation}
    \tau(z) = \int_0^z [1 - x_{\rm{HI}}(z')] n_e(z') \sigma_T c\, dt \ ,
\end{equation}
where the Thomson cross-section $\sigma_T = 6.65 \times 10^{-25}$ cm$^2$, $x_{\rm{HI}}$ is the neutral hydrogen fraction, and $c\, dt$ is the differential physical distance (in terms of the physical time $t$). The electron number density ($n_e$) accounts for contributions from ionized hydrogen and helium, and it can be written as:

\begin{equation}
    n_e(z) = \Omega_b (1 - Y_{\rm{He}}) \frac{\rho_{\rm{cr}}}{m_{\rm{H}}} (1 + z)^3 \left(1 + p \frac{n_{\rm{He}}}{n_{\rm{H}}}\right)\ ,
\end{equation}
where $\rho_{\rm{cr}}$ is the present critical density of the Universe, $\Omega_b$ is the baryon density parameter, $m_{\rm{H}}$ is the mass of hydrogen, $Y_{\rm{He}}$ is the helium mass fraction, and $n_{\rm{He}}/n_{\rm{H}}$ the helium-to-hydrogen ratio. In terms of the fraction of helium nuclei $f_{\rm{He}}$, $Y_{\rm{He}} = 4f_{\rm{He}}/(1- f_{\rm{He}} + 4f_{\rm{He}})$ and $n_{\rm{He}}/n_{\rm{H}} = \frac{f_{\rm{He}}}{1-f_{\rm{He}}}$. We assume 
$f_{\rm{He}} = 0.0732$ (corresponding to $Y_{\rm{He}}=24\%$). The electron density is modified with redshift to account for the small effect of helium reionization. For $ z < 3 $, we assume that helium is doubly ionized ($p=2$), while for $ z \geq 3 $ it is singly ionized ($p=1$).

We normalize our models to the total optical depth to the CMB. We show our results for the following two cases:
\begin{enumerate}
    \item ``Standard": $\tau=0.054$, the value favored by Planck measurements \citep{Planck2020}.
    \item ``High $\tau$": $\tau=0.068$, a relatively high example value, namely the standard value $+\ 2\sigma$ [see also \citet{2025arXiv250416932S}].
\end{enumerate}

\subsection{Modified 21-cm brightness temperature due to scattering}\label{sec:observed_T21}

The analysis is similar to that of CMB photons from the surface of last scattering. In our case, as CMB photons travel from redshift $ z_0 $ (where they reached the 21-cm wavelength and interacted with hydrogen atoms) to $ z = 0 $, they interact with ionized regions at intermediate redshifts $ z $ (where the wavelength is $21(1+z_0)/(1+z)$~cm). The optical depth ($\tau$) quantifies the extent to which these photons were re-scattered by the medium. At each infinitesimal redshift interval $dz$ around redshift $z$, the incremental optical depth $ d\tau $ represents the probability of a photon being scattered or absorbed. This $ d\tau $ corresponds to the fraction of light scattered out of the line of sight, with an equivalent amount of light scattered into the line of sight from other directions. The latter corresponds to the local CMB sky at a wavelength of $21(1+z_0)/(1+z)$~cm and thus averages over photons from all directions that hit 21~cm at the same redshift $z_0$; their average intensity thus reflects the mean 21-cm intensity at redshift $z_0$. 

Suppose the 21-cm brightness temperature at redshift $z_0$ in some direction (labeled $x$) is $T_{{21,x}}$, while the 21-cm global mean brightness temperature across the sky at that redshift is $T_{21,m}$. The observed 21-cm brightness temperature is a combination of two contributions:
\begin{enumerate}
    \item The fraction of the original light from $z_0$ that is unscattered: $e^{-\tau(z_0)}$. This contributes $T_{21,x}\ e^{-\tau(z_0)}$, 
    \item The fraction of light that is scattered into the line of sight, which had the mean brightness temperature: $1-e^{-\tau(z_0)}$. This contributes $T_{21,m}[1-e^{-\tau(z_0)}]$. 
\end{enumerate}
Thus, the total observed brightness temperature is given by 
\begin{equation}
    T_{\rm{21, obs}} = T_{21,x}\ e^{-\tau(z_0)} + T_{21,m}(1-e^{-\tau(z_0)}) \ .
\end{equation} Since interferometers measure the 21-cm brightness temperature relative to the mean at each redshift, the observed brightness temperature fluctuation is:
\begin{equation}
    \delta T_{21, \rm{obs}} \equiv T_{\rm{21, obs}} - T_{\rm{21,m}} = (T_{21,x} - T_{\rm{21,m}})e^{-\tau(z_0)} \ .
\end{equation}

This shows that the original fluctuations are attenuated by the factor $e^{-\tau(z_0)}$ due to the scattering. This is directly applicable to 21-cm images. When the fluctuations in the brightness temperature are measured statistically, the common statistic is the power spectrum. It measures the variance of the fluctuations, $P(k) \propto \langle|\delta T_{21, \rm{obs}}|^2 \rangle$, and so is suppressed by a factor of $e^{-2\tau(z_0)}$.

\section{Results}

\begin{figure*}
\centering
\includegraphics[scale=0.5]{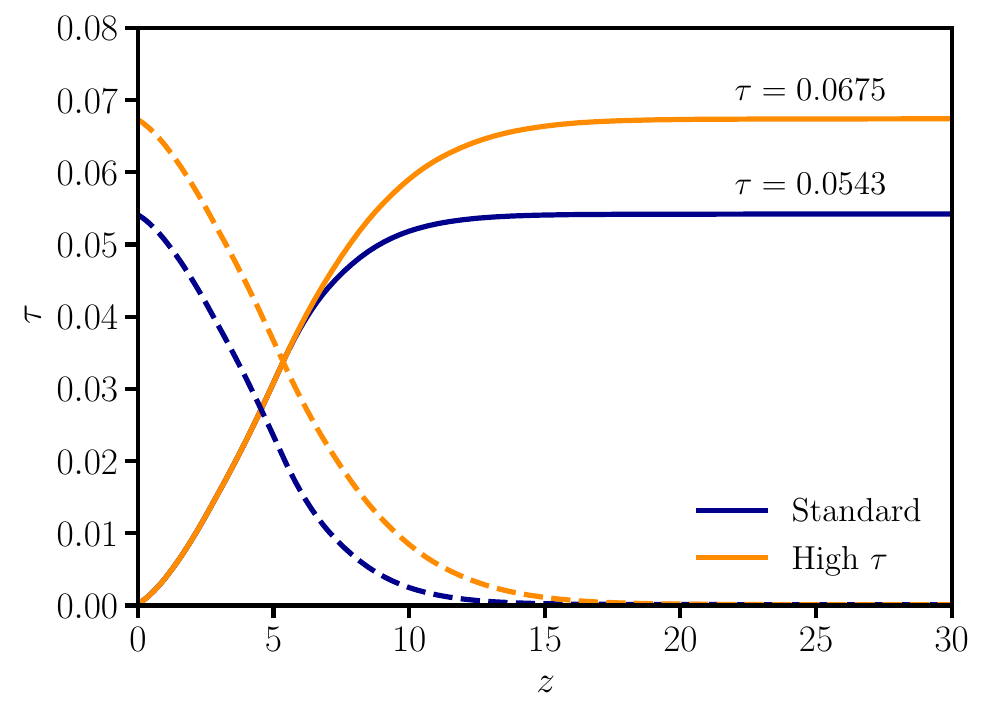}
\includegraphics[scale=0.5]{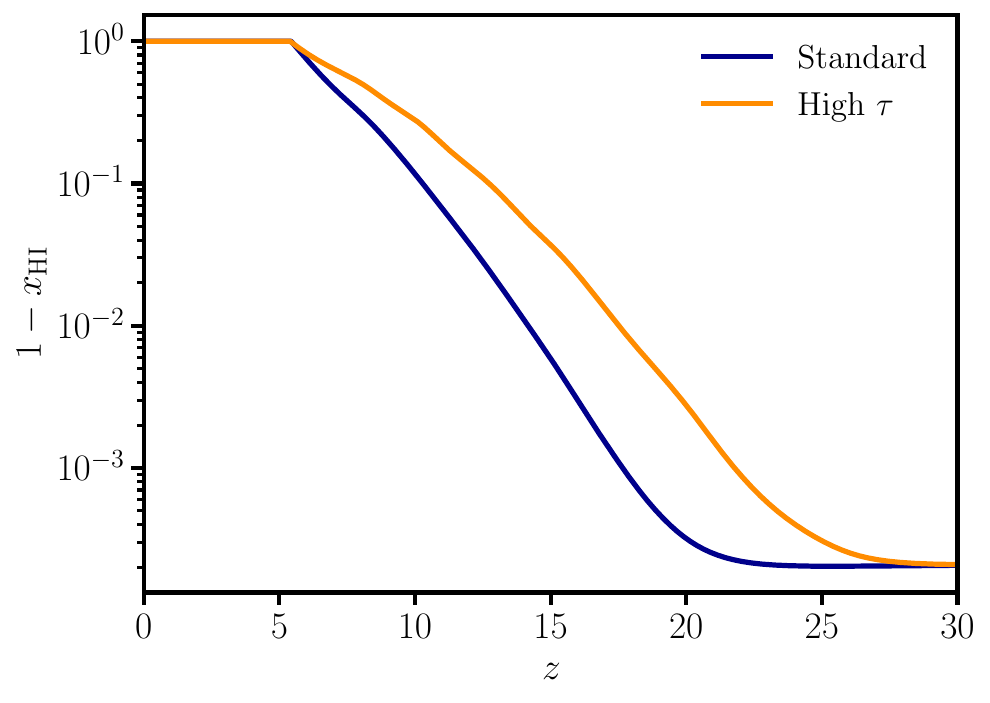}
\caption{\textbf{Left panel:} Cumulative optical depth ($\tau$) to the CMB as a function of $z$ for our ``Standard" and ``High $\tau$" case.  Here The solid curves represent the total integrated optical depth from $z=0$ to $z$ for the two cases considered: the ``Standard" case (blue) yielding $\tau = 0.0542$ at $z=30$, and the ``High $\tau$" case (orange) yielding $\tau = 0.0674$. The corresponding dotted curves illustrate the residual contribution to the total $\tau$ from redshifts greater than $z$, emphasizing the dominance of high-$z$ contributions in both scenarios. \textbf{Right panel:} Ionized hydrogen fraction ($1 -x_{\rm{HI}}$) as a function of $z$ for the ``Standard" and ``High $\tau$" cases.}\label{fig:tau_xHI_vs_z}
\end{figure*}

We first set the stage by showing (Fig.~\ref{fig:tau_xHI_vs_z}, right panel) the progress of reionization, i.e., the evolution of the ionized hydrogen fraction, derived from the semi-numerical simulations for the ``Standard" or ``High $\tau$" cases. Both models complete reionization at $z$ just below 6 (where we slightly extrapolated beyond our simulation end at $z=6$). At the opposite end, at the highest redshifts, a residual floor in the ionized fraction emerges, attributable to the small non-zero ionized fraction persisting after recombination. Between $z \approx 6$ and $z \approx 28$, the ``Standard" case has a lower ionized fraction compared to the ``High $\tau$" case. For instance, at $z=10$ the ionized fraction is 0.118 (``Standard") or 0.282 (``High); at $z = 20$, these values drop to $3.2 \times 10^{-4}$ and $2.75 \times 10^{-3}$, respectively. 

The left panel of Fig. \ref{fig:tau_xHI_vs_z} illustrates the evolution of the cumulative optical depth $\tau (z)$ to the CMB for the ``Standard" and ``High $\tau$" cases. Here, solid curves represent the total integrated optical depth from redshift 0 to $z$ with the "Standard" case (blue) reaching $\tau = 0.0543$ and the "High $\tau$" case (orange) reaching $\tau = 0.0675$ by $z = 30$. Dotted curves complement these by showing the contribution to the total $\tau$ from redshifts greater than $z$. The dotted curves emphasize that a significant part of the optical depth originates from the EoR ($z \approx 6$ to 15). For instance, in the ``Standard" case, $\tau = 0.0385$ at $z=6$, indicating that $29\%$ of the total $\tau$ arises from $z>6$, whereas in the ``High $\tau$" case, $\tau=0.0388$ at $z=6$, so that $43\%$ of the total $\tau$ comes from $z>6$. Values of $\tau$ at additional redshifts are listed in 
Table~\ref{tab:table2}. 

\begin{table}
\centering
\begin{tabular}{lcc} 
\hline
$z$         & Standard & High $\tau$  \\ 
\hline\hline
3  & 0.0158 & 0.0158 \\
6  & 0.0385 & 0.0388 \\
10 & 0.0519 & 0.0591 \\
15 & 0.0541 & 0.0664 \\
30 & 0.0542 & 0.0674 \\
$\infty$ & 0.0543 & 0.0675 \\
\hline
\end{tabular}
\caption{A few illustrative values of $\tau$, at $z=3, 6, 10, 15, 30$, and the total value. These values correspond to the solid curves in the left panel of Fig.~\ref{fig:tau_xHI_vs_z}.}
\label{tab:table2}
\end{table}

\begin{figure}
\centering
\includegraphics[scale=0.5]{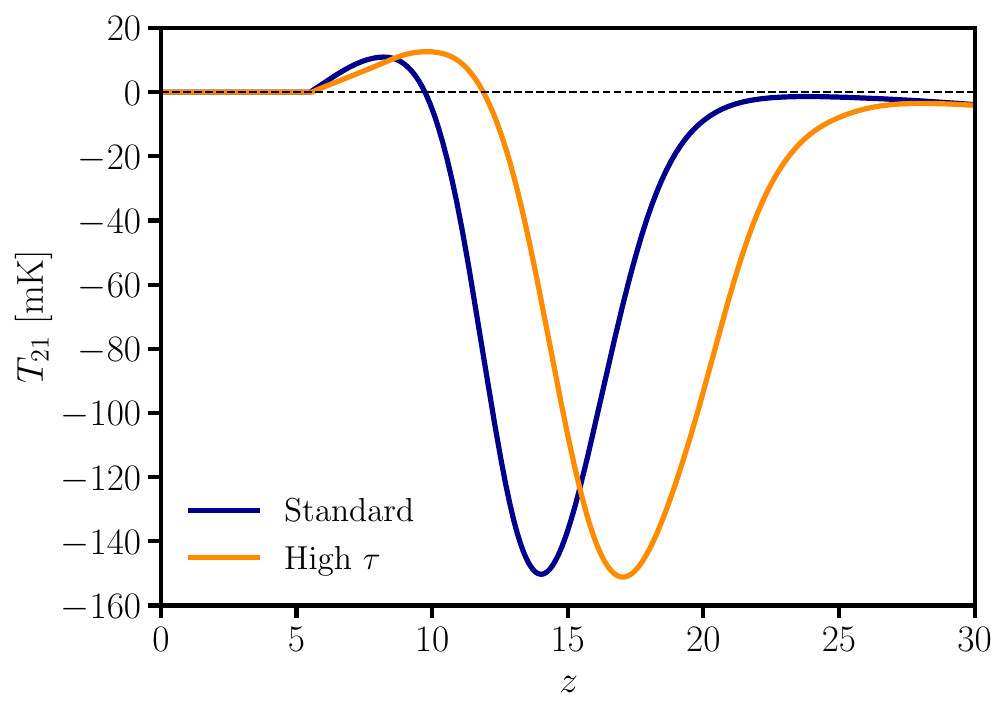}
\caption{Global signal as a function of $z$ for our ``Standard" and ``High $\tau$" cases. Also shown is $T_{21}=0$ (dashed black line).}\label{fig:global_signal_vs_z}
\end{figure}

Fig. \ref{fig:global_signal_vs_z} presents the global 21-cm signal as a function of $z$ for the ``Standard" and ``High $\tau$" cases. Since our ``High $\tau$" case corresponds to a lower minimum circular velocity ($V_{\rm{C}}$) for star formation and a higher reionization efficiency $\zeta$  compared to the ``Standard" case, the absorption trough occurs at an earlier cosmic time (higher redshift, $z \sim 17$), whereas the ``Standard" case reaches its trough at $z \sim 14$. Despite this temporal shift, the amplitude of the absorption troughs are comparable, approximately $-150$ mK. By $z \sim 5.5$, the global signal diminishes to zero for both models, marking the transition to a fully ionized universe. 

\begin{figure*}
\centering
\includegraphics[scale=0.5]{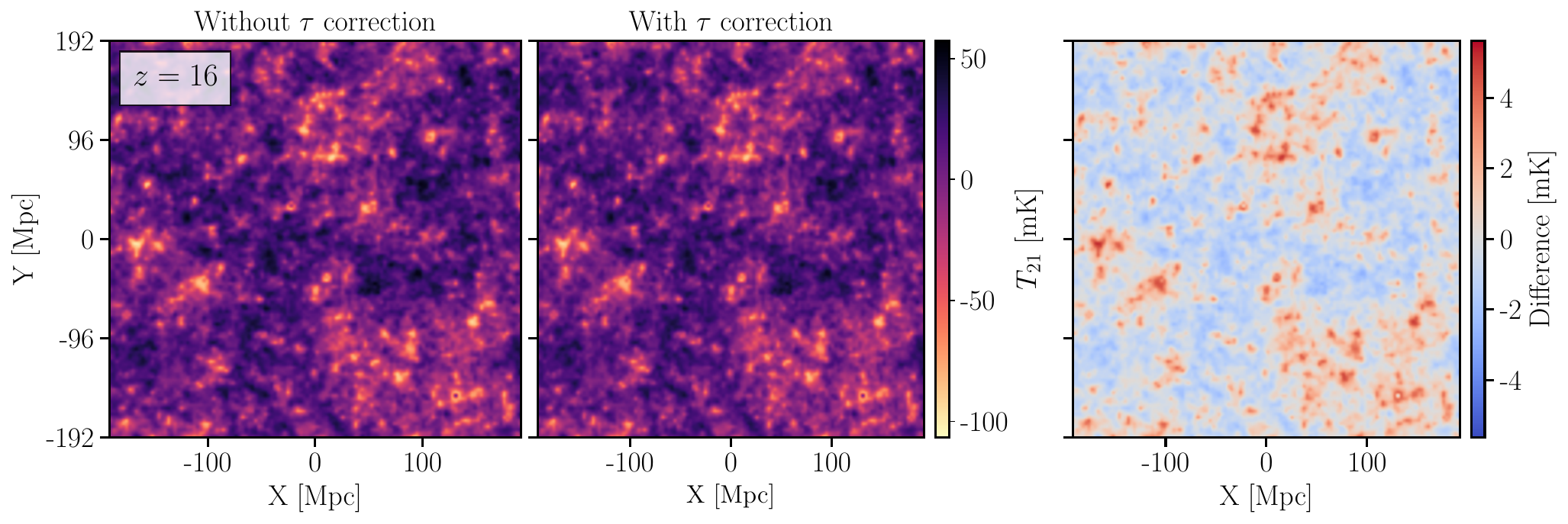}
\caption{21-cm slices at $z=16$ without (\textbf{left panel} ) or with (\textbf{middle panel} ) the $\tau$ correction for the ``Standard" model. As appropriate for observations with an interferometer, each slice is shown relative to the mean intensity (global signal) at that redshift which is -103.9~mK. \textbf{Right panel:} The difference between the two slices (With $\tau$ minus Without $\tau$). Note that the pixels in these maps (as in our simulations) are 3~Mpc on a side.}\label{fig:21cm_slices}
\end{figure*}

Moving on to our main results, Fig. \ref{fig:21cm_slices} presents 21-cm brightness temperature slices at $z=16$ for the ``Standard" case, comparing maps without or with the $\tau$-correction, alongside the difference induced by this correction. Since with the $\tau$ correction, the original 21-cm fluctuations are attenuated by $e^{-\tau(z)}$, the brightest regions (i.e., those with more strongly negative brightness temperatures) in the 21-cm map result in positive differences. While the differences between the slices are difficult to see visually, these differences (of up to 5.6~mK in this example) are important for precise predictions. 

\begin{figure*}
\centering
\includegraphics[scale=0.5]{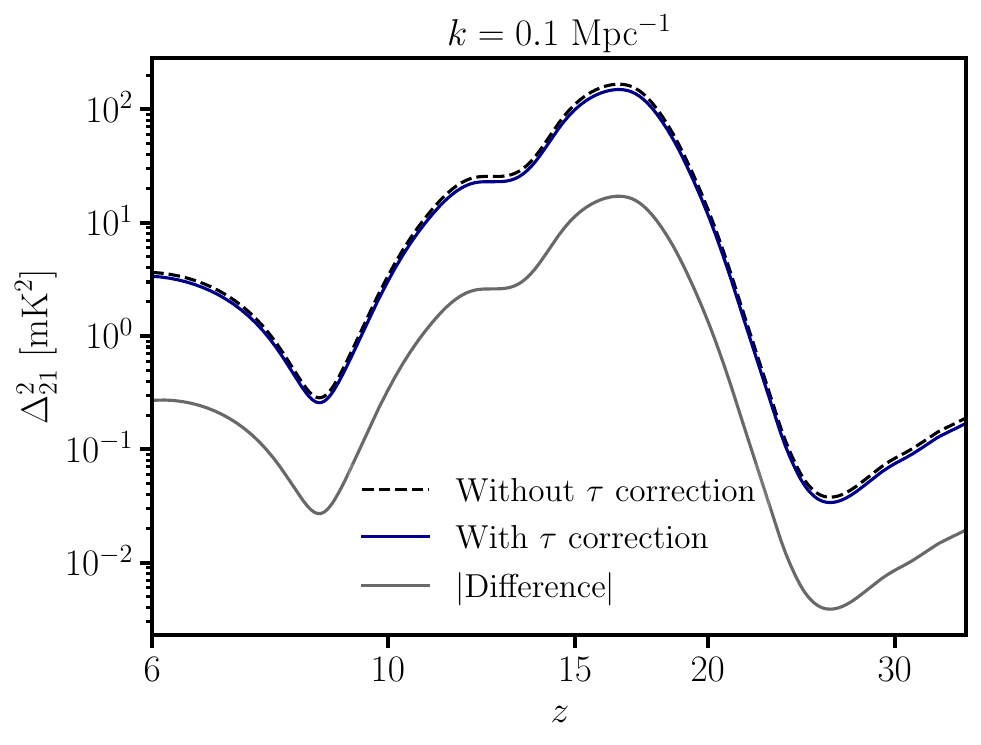}
\includegraphics[scale=0.5]{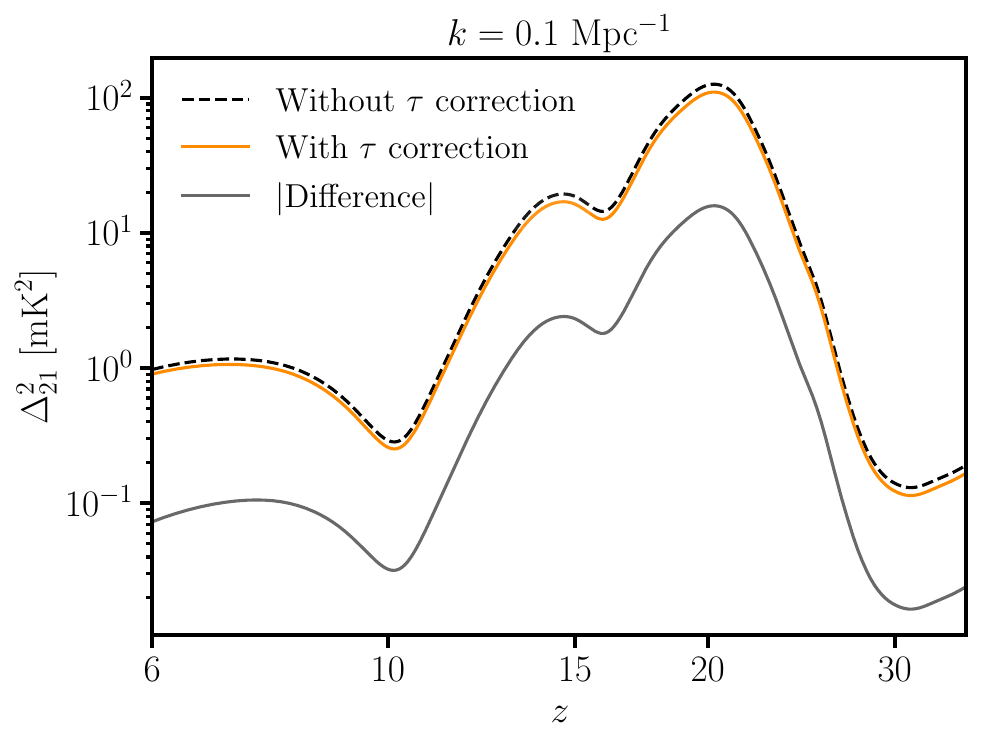}
\caption{21-cm power spectrum as a function of $z$ at $k=0.1$ Mpc$^{-1}$, without or with the $\tau$ correction. \textbf{Left panel:} ``Standard" model. \textbf{Right panel:} ``High $\tau$" model. Dashed curves show the power spectrum without the $\tau$ correction while solid gray curved give the absolute difference between the $\tau$ corrected and uncorrected cases, revealing differences of $\sim 10\%$ throughout reionization and cosmic dawn.}\label{fig:21cm_ps}
\end{figure*}

In Fig. \ref{fig:21cm_ps} we illustrate the effect of the $\tau$ correction on the 21-cm power spectrum, shown as a function of $z$, for the ``Standard" (left panel) or ``High $\tau$" (right panel) cases. We show the wavenumber $k=0.1$ Mpc$^{-1}$ but the relative effect of the $\tau$ correction is independent of scale. Each panel shows the (absolute) difference between the $\tau$-corrected and uncorrected power spectra (solid gray curve). As discussed in section~\ref{sec:observed_T21}, the scattering of the 21-cm/CMB photons with electrons attenuates the 21-cm fluctuations by a factor of $e^{-\tau(z)}$, suppressing the power spectrum by $e^{-2\tau(z)}$, which leads to a noticeable reduction. In general, the $\tau$ correction is essential for accurate comparisons with observations from experiments like NenuFar \citep{NenuFAR}, LOFAR \citep{LOFAR}, HERA \citep{HERA}, the MWA \citep{MWA}, and the Square Kilometre Array (SKA) \citep{SKA}. 

\begin{figure}
\centering
\includegraphics[scale=0.5]{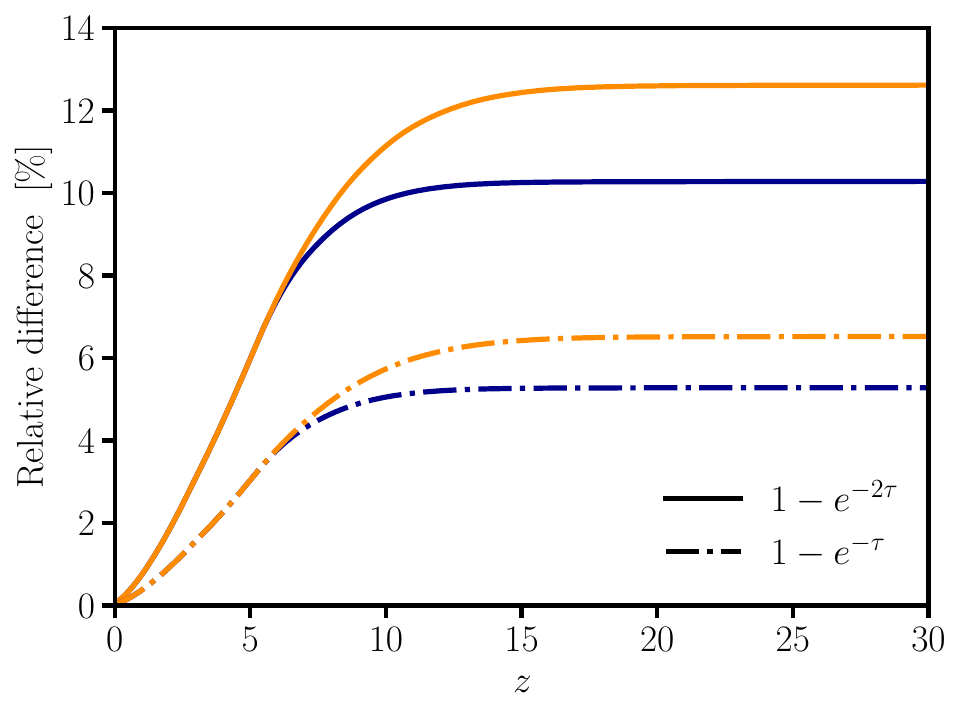}
\caption{Relative difference (in percent) as a function of $z$, for the ``Standard" (blue curves) and ``High $\tau$" (orange curves) cases. The dashed curves represent the imaging factor $(1-e^{-\tau})$, which quantifies the suppression of the mean 21 cm brightness temperature. The solid curves represent the power spectrum factor $(1-e^{-2\tau})$, which quantifies the suppression of the 21-cm power spectrum. Here we calculate the relative differences directly from the calculated $\tau$. These factors increase with $z$, reaching up to $\sim 10.3\%$ for the ``Standard" case and $\sim 12.6\%$ for the ``High $\tau$" case.}\label{fig:relative_diff}
\end{figure}

Fig. \ref{fig:relative_diff} summarizes the key quantitative measure of the impact of Thomson scattering on 21-cm observables, derived directly from the $\tau(z)$ curves shown in the left panel of Fig.~\ref{fig:tau_xHI_vs_z}. For the 21-cm map, the effect reaches up to $5.3\%$ for the ``Standard" case and $6.5\%$ for the ``High $\tau$" case, while for the power spectrum, it reaches up to $10.3\%$ for the ``Standard" case and $12.6\%$ for the ``High $\tau$" case. The minimum correction during the EoR, at $z=6$, is $3.8\%$ for images and $7.4\%$ for the power spectrum. These results show that, although the $\tau$-correction remains modest, its effect is non-negligible during cosmic dawn and the epoch of reionization, skewing 21-cm power spectrum estimates by $\sim 10\%$ if overlooked, underscoring its importance for precise cosmological analyses. 

\section{Summary \& Discussion}

%% The "ht!" tells LaTeX to put the figure "here" first, at the "top" next
%% and to override the normal way of calculating a float position.
%% The asterisk after "figure" tells the compiler to span multiple columns
%% if a two column style is selected.

We have shown that Thomson scattering of 21-cm/CMB photons affects 21-cm cosmology similarly to its well-known effect on CMB anisotropies. Thus, this scattering suppresses the fluctuations in 21-cm images as well as statistical measures of the fluctuations, including the 21-cm power spectrum. A difference is that in the case of primary CMB anisotropies, the suppression depends only on a single number, namely the total optical depth $\tau$ out to high redshifts (although the effect of reionization on CMB polarization does depend on the detailed reionization history, e.g., \citet{2021PhRvD.104f3505H}). In the 21-cm case, the effect accumulates during the epoch of reionization, which is where many 21-cm observations focus, and thus its redshift dependence plays a major role. 

The effect can be calculated precisely at low redshifts, below the EoR, though we do not currently know the precise redshift of the end of reionization, and there is also up to a few percent relative uncertainty due to the small fraction of neutral hydrogen that remains in galaxies post-reionization (whose effect we have neglected in this paper). We note that the $\tau$ effect also similarly influences intensity mapping at low redshifts (e.g., when the 21-cm line is used for measuring galaxy clustering due to the neutral hydrogen content of galaxies), but, e.g., the effect on the power spectrum at $z=3$ is a suppression by only 3.1\%. During the EoR, the effect starts (e.g., at $z=6$) at $3.8\%$ for images and $7.4\%$ for the power spectrum, and then rises up to $5.3\%$ for images and $10.3\%$ for the power spectrum (in our ``Standard" case), or $6.5\%$ for images and $12.6\%$ for the power spectrum (in our ``High $\tau$" case). The effect is expected to saturate to near its final $z \rightarrow \infty$ value by $z \sim 12 - 15$ depending on the reionization history.

While we have calculated the main effect of Thomson scattering on 21-cm observations, we did make some simplifying assumptions. For example, we assumed that the mean intensity of the photons scattering into our direction from others reflects the global mean intensity at the original redshift. This is accurate if the photons come from large distances, but at short distances they reflect local conditions rather than the global mean. However, short distances contribute little to the overall $\tau$. To estimate an extreme case, consider a 50~Mpc (comoving) line of sight through an ionized region. At $z \sim 6$ this contributes 0.1\% to $\tau$, so should cause such a small, additional fluctuation relative to a 50~Mpc line of sight through a still neutral region. This rises to 0.26\% at $z=10$, though such large fluctuations are less likely to be present on that scale at that redshift. This effect would act differently on the sky (at a given redshift) compared to the line of sight (where the redshift changes). This introduces a small new line of sight anisotropy. Even if we neglect the small effect just discussed, so that we treat $\tau$ as uniform on the sky at a given redshift, along the line of sight it will still reduce fluctuations more strongly at higher redshifts, causing an anisotropy. Again, on typical scales at which 21-cm fluctuations are observable, the size of this effect should be limited to a fraction of a percent.

We have illustrated the range of the possible effect using two different astrophysical models that were normalized based on the measured optical depth to the CMB and its uncertainty. We note, though, that current parameter tensions in cosmology remain unresolved, and imply that the total optical depth to the CMB could still possibly be significantly higher than even our ``High $\tau$" case \citep{2025arXiv250416932S}. For instance, a $\tau$ value of 9\% would imply a power spectrum suppression of 16.5\%. More generally, the $\tau$ effect should now be included in all 21-cm predictions from models that predict a reionization history, whether derived from analytical models or semi-numerical or full numerical simulations.

%% Please use the acknowledgment and contribution environments. This will 
%% be anonomyized when the "anonymous" style option is used. 
\begin{acknowledgments}
SS and RB acknowledge the support of the Israel Science Foundation (grant no.\ 1078/24).
\end{acknowledgments}

\software{\texttt{Numpy} \citep{harris2020array}, \texttt{Scipy} \citep{2020SciPy-NMeth}, \texttt{matplotlib} \citep{Hunter:2007}}


%\appendix

\bibliography{sample701}{}
\bibliographystyle{aasjournalv7}



\end{document}